\documentclass[aps,prb,twocolumn,showpacs,showkeys,superscriptaddress]{revtex4-1}

\usepackage[pdftex]{graphicx}
\usepackage{bm}

\begin{document}
\title{Fraunhofer patterns for    Josephson junctions in narrow  thin-films with  vortices trapped in one of the banks}

\author{V. G. Kogan}
\affiliation{Ames Laboratory, US Department of Energy, Ames, Iowa 50011, USA }
\author{ R. G. Mints }
\affiliation{The Raymond and Beverly Sackler School of Physics and Astronomy, Tel Aviv University, Tel Aviv 69978, Israel }
\date{\today}

 \begin{abstract}
It is shown that a vortex  trapped in one of the banks of a planar edge-type Josephson junction in a narrow thin-film  superconducting strip
can change drastically the field dependence of the junction critical current $I_c(H)$. When the vortex is trapped at certain positions in the strip middle, the  pattern $I_c(H)$ has zero at $H=0$ instead of the traditional maximum of '0-type' junctions. The number of these positions is equal to the number of vortices trapped at the same location.
When the junction-vortex separation exceeds approximately $2W$,   $I_c(H)$ is no longer sensitive to the vortex presence.
\end{abstract}

\pacs{74.60. Ec, 74.60. Ge}

\maketitle

\section{ Introduction}

The very fact that  Abrikosov vortices in the vicinity of Josephson junctions affect the junction properties  is well documented and not surprising since the phase associated with vortex affects the junction phase difference. \cite{Ustinov,Fistul}
Recent experiments with a vortex trapped in one of the banks of an edge-type  planar junction in a thin-film superconducting strip  showed that the vortex causes an extra phase difference on the junction that depends on the vortex position. \cite{Krasnov}   The effect is strong in particular when the vortex is close to the junction, the situation when  the junction behavior is changed from the conventional ``zero"-type  to that of the $\pi$-junction.

Here we study how the field dependence of maximum critical tunneling currents $I_c(H)$, commonly called the Fraunhofer pattern, changes when a vortex trapped in one of the junction thin-film banks changes its position.
In principle, this effect can be utilized  for manipulating Josephson currents by  controlling  the   vortex position.

%%%%%
\section{ Approach}
%%%%%%%

\begin{figure}[h]
\begin{center}
 \includegraphics[width=5.cm] {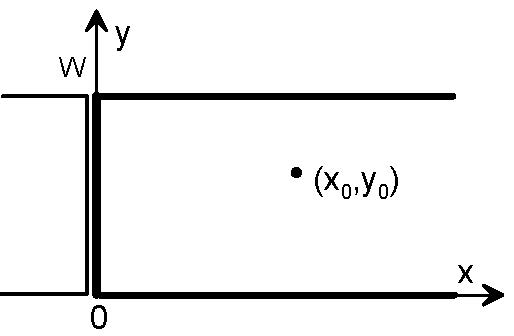}
\caption{  The superconducting thin-film strip with a Josephson junction at $x=0$   and a vortex at $(x_0,y_0)$.  The half-strip containing the vortex is outlined by   thick lines.
}
\label{f0}
\end{center}
\end{figure}

 Consider a  thin-film strip of a width $W$ with an edge-type  Josephson junction across the strip which cuts the strip in two half-strips, Fig.\,\ref{f0}. The strip is narrow: $W \ll \Lambda=2\lambda^2/d$  where $\lambda$ is the London penetration depth of the film material and $d$ is the film thickness. Choose $x$ along the strip and $y$ across so that $0<y<W$ and the junction is at $x=0$. Let a vortex be trapped at some point $\bm r_0=(x_0,y_0)$ in the right half-strip ($x_0>0$).

The London equation integrated  over the film thickness  for the half-strip with vortex (shown by a thick line in Fig.\,\ref{f0}) is:
\begin{equation}
h_z + {2\pi\Lambda\over c}\,{\rm curl}_z\,{\bm g}= \phi_0\delta(\bm r-\bm r_0)\,.
\label{London}
\end{equation}
Here   ${\bm g}$ is  the sheet current density and $h_z$ consists of the applied field $H$ and the self-field of the current $\bm g$.

The self-field of the current $\bm g$ is of the order $g/c$, whereas the second term on the left-hand side of
Eq.\,(\ref{London}) is of the order $g\Lambda/cW\gg g/c$. Hence, in narrow strips with $W\ll \Lambda$, the self-field can be disregarded, unlike the {\it applied} field $H\hat{\bm z}$. Introducing the scalar stream function $S$ via $\bm g = {\rm curl}_z S(x,y)\hat z$, we obtain instead of Eq.\,(\ref{London}):
\begin{equation}
\nabla^2 S =  \frac{cH}{2\pi\Lambda}-\frac{c\phi_0}{2\pi\Lambda}\delta(\bm r-\bm r_0)\,\,.
\label{eq_02}
\end{equation}
This is a {\it linear} Poisson equation,    that  formally simplifies the problem as compared to the general Eq.\,(\ref{London}). Physically, this simplification comes about since in narrow films the major contribution to the system energy is the kinetic energy of supercurrents, while  their magnetic energy can be disregarded.

 The boundary condition $g_y=0$ at the strip edges translates to $S=0$ at the edges $y=0,W$ (in the absence of transport current). Besides,   one can disregard Josephson tunneling currents relative to those of the vortex, i.e. to set $g_x(0,y)=0$ as well.  The Green's function
$G({\bm r},{\bm r^\prime})$ which satisfies $ \nabla^2 G =-4\pi
\delta ({\bm r}-{\bm r^\prime})$ (as the electrostatic potential of a unit  linear charge  at $\bm r^\prime$) with zero boundary conditions at the  edges of the half-srtip (deleneated in Fig.\,\ref{f0} by  thick lines) is found by conformal mapping: \cite{Morse,Clem,last}
 \begin{eqnarray}
u+i v=-i\cosh \pi(x+i y)
 \label{eq3}
\end{eqnarray}
 transforms the   half-plane $u>0$ to the half-strip of a width 1 (hereafter we use $W$ as a unit length).  Explicitly, this transformation  reads:
 \begin{eqnarray}
u =\sinh \pi x\,\sin \pi y  \,, \qquad  v=-\cosh \pi x\,\cos \pi y\,.
 \label{eq4}
\end{eqnarray}

  The complex potential $G(w,w^\prime)$ for a linear unit charge  at $w^\prime=u^\prime+iv^\prime$ at the half plane $ u>0 $  is: \cite{LL}
 \begin{eqnarray}
G= -2  \ln\frac{w-w^\prime}{w-\tilde w^\prime} = -2 \left[\ln\frac{r_1}{r_2}+i (\theta_1-\theta_2)\right]
 \label{eq6}
\end{eqnarray}
where $w=u+iv$, $ \tilde w^\prime=-u ^\prime+i\,v^\prime$ is the position of fictitious image source on the opposite side of the grounded plane  $u=0$.   The corresponding moduli and phases are:
  \begin{eqnarray}
r_1 &=&\sqrt{(u-u^\prime)^2+(v-v^\prime)^2}  \,, \nonumber\\
r_2&=&\sqrt{(u+u^\prime)^2+(v -v^\prime)^2}  \,,  \nonumber\\
 \theta_1 - \theta_2&=&\tan^{-1}\frac{v-v^\prime}{u-u^\prime} -  \tan^{-1}\frac{v-v^\prime}{u+u^\prime} \,.
 \label{eq8}
\end{eqnarray}
Below we  evaluate the phase at the junction bank $x=+0$ and make use   of
\begin{eqnarray}
  {\rm Im}G(0,y;\bm r^\prime) = -
 4 \tan^{-1}\frac{\cos \pi y-\cosh \pi x^\prime\,\cos \pi y^\prime}{\sinh \pi x^\prime\,\sin \pi y^\prime}
 \,,\qquad \label{eq_12b}
\end{eqnarray}
which follows from Eqs.\,(\ref{eq6}) and (\ref{eq8}).

 We now note that the sheet current is expressed either in terms of the gauge invariant phase $\varphi$ or via the stream function $S$:
${\bm g}=-(c\phi_0/4\pi^2\Lambda)\nabla \varphi
={\rm curl}\,S {\bm z}$. (This  relation written in components shows that
$(4\pi^2\Lambda/c\phi_0)S({\bm r})$ and $\varphi({\bm r})$
are the real and imaginary parts of an analytic function.)
In particular, we have
\begin{eqnarray}
\frac{ \partial\varphi}{\partial y}  = -\frac{4\pi^2 \Lambda}{c \phi_0}\,g_y\,.
\label{new7}
\end{eqnarray}
On the other hand, the sheet current $\bm g_0$ for a {\it unit} $\delta$-function source   can be expressed either via  real or imaginary parts  of  $G$. \cite{LL}  In particular, we have:
  \begin{eqnarray}
g_{0y}= -\frac{\partial \,{\rm Re}G}{\partial x}=-\frac{\partial\, {\rm Im}G}{\partial y}  .
 \label{g0}
\end{eqnarray}

 %%%%%%%%
\subsection {Contribution of the  field $\bm H$ at the right bank to the phase difference at the junction}
%%%%%%%%

\begin{figure}[h]
\begin{center}
 \includegraphics[width=7cm] {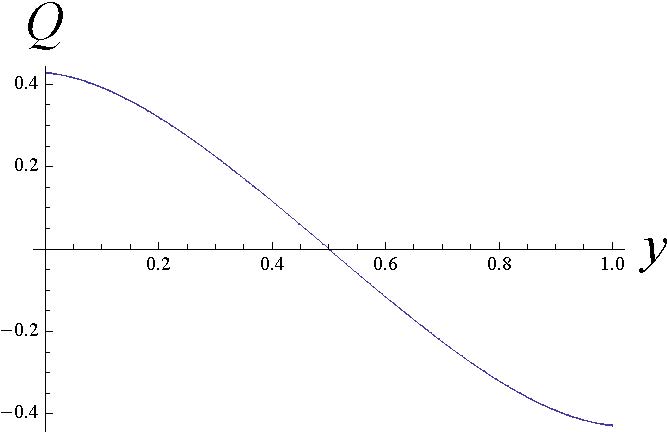}
\caption{(Color online)  The integral $Q$ of Eq.\,(\ref{eq_12c})    vs $y$ for the strip in magnetic field calculated numerically and superimposed with the expression $0.43 \cos \pi y -0.03\sin 2\pi y$. Numerical check shows that the difference of the two is less than 0.5\%.
}
\label{f1}
\end{center}
\end{figure}

The solution of Eq.\,(\ref{eq_02}) without a vortex, $\nabla^2 S = -4\pi(-  cH/8\pi^2\Lambda)$, is
\begin{eqnarray}
&&{\rm Im}S(0,y)=-W^2\int d\bm r^\prime \frac{cH}{8\pi^2\Lambda} {\rm Im}G(0,y;{\bm r^\prime})\nonumber\\
&&=\frac{cHW^2}{2\pi^2\Lambda}\int  d{\bm r}^\prime   \tan^{-1}\frac{\cos \pi y-\cosh \pi x^\prime\,\cos \pi y^\prime}{\sinh \pi x^\prime\,\sin \pi y^\prime}
 \,,\qquad
\label{eq_12c}
\end{eqnarray}
 where the  integrals are extended over the half-strip: $0<x^\prime<\infty$, $0<y^\prime<1$. The last integral $Q$ can be evaluated  in terms of Lerch transcendents,  \cite{Clem}   which are not particularly illuminating. Hence, for each $y$ we do the integration  numerically.  The result is shown in Fig.\,\ref{f1}. The function  $Q(y)$ can be approximated as
 \begin{eqnarray}
Q\approx 0.43 \cos \pi y -0.03\sin 2\pi y
 \label{Q(y)}
\end{eqnarray}
 with accuracy less than 0.5\%. In fact, the numerically evaluated $Q$ and this approximation cannot be distinguished at Fig.\,\ref{f1}. The quantity $Q$ has also been  calculated  employing a different method in Ref.\,\onlinecite{Maayan}. We use the approximation (\ref{Q(y)}) in the numerical work below.

At the junction bank $x=+0$, $g_y(0,y)=-\partial_y {\rm Im}S(0,y)$, and we obtain with the help of Eqs.\,(\ref{new7}), (\ref{eq_12c}), and (\ref{Q(y)}) :
\begin{eqnarray}
\frac{ \partial\varphi}{\partial y}  = -\frac{h}{2}\,\frac{ \partial Q}{\partial y}\,.
 \label{new12}
\end{eqnarray}
or after integration over $y$:
\begin{eqnarray}
 \varphi_H(+0, y)= -\frac{h}{2}\, Q(y) +\varphi_0\,,\qquad h=\frac{4W^2H}{\phi_0}\,.
 \label{phiH/2}
\end{eqnarray}
The subscript $H$ here is to indicate that this contribution to the phase is due to the applied field; $\varphi_0$ is an arbitrary constant. \cite{Maayan} \\

%%%%%%%%
\subsection {Contribution of a vortex at $\bm r_0$ to the phase difference at the junction.}
%%%%%%%%

To find this contribution, we use the relation (\ref{new7})   along with
\begin{eqnarray}
g_y(+0,y) = -\frac{\partial\, {\rm Im}S }{\partial y} = -
\frac{c \phi_0}{8\pi^2 \Lambda}\,\frac{\partial\, {\rm Im}G(0,y;\bm r_0)}{\partial y}.\qquad
\label{gy-phi-vort}
\end{eqnarray}
We obtain after integration over $y$:
\begin{eqnarray}
\varphi_v( y;\bm r_0) &=& \frac{1}{2}\,{\rm Im}G(0,y;\bm r_0) \nonumber\\
&=&- 2  \tan^{-1}\frac{\cos \pi y-\cosh \pi x_0\,\cos \pi y_0}{\sinh \pi x_0\,\sin \pi y_0} ,\qquad
\label{phi-vort}
\end{eqnarray}
where an arbitrary constant  is omitted.

It is worth observing that at large vortex-junction separations  $x_0\gg 1$, this contribution is a constant which does not depend on $x_0$:
\begin{eqnarray}
\varphi_v( y;\bm r_0) = \pi( 2y_0-1) +{\cal O}(e^{-\pi x_0});
\label{C}
\end{eqnarray}
in other words, corrections to this constant are exponentially small with the length scale $W/\pi$.\\

%%%%%%%%
\subsection  {The critical current $\bm{I_c(H,\bm r_0)}$}
%%%%%%%%

The total phase difference at the junction is
\begin{eqnarray}
\varphi(y;H,\bm r_0 )= \varphi_H(y)+\varphi_v( y;\bm r_0) + \varphi_0 \,.
\label{phasediff}
\end{eqnarray}
The field induced phase difference $\varphi_H(y)$ is twice as large as $\varphi_H(+0,y)$ which was evaluated for a half-strip in Eq.\,(\ref{phiH/2}) because both right and left half-strips contribute equally.

 The Josephson current density  $ g_c\sin\varphi(y ) $    integrated over the junction length gives the total  current $I $:
\begin{eqnarray}
&&\frac{I ( H,\bm r_0)}{ g_cW} =A \cos\varphi_0  +B \sin\varphi_0\,,  \label{I}\\
 &&A= \int_0^1 \sin(\varphi_H +\varphi_v )  dy,\quad B= \int_0^1 \cos(\varphi_H +\varphi_v )  dy.\qquad
\nonumber
\end{eqnarray}
The right-hand side of Eq.\,(\ref{I})  is easily transformed to
\begin{eqnarray}
\sqrt{A^2+B^2}\cos (  \varphi_0 -\psi)\,, \quad \psi =\sin^{-1}\frac{B}{\sqrt{A^2+B^2} } .\qquad
\label{psi}
\end{eqnarray}
 Maximizing this relative to the free parameter $\varphi_0$ one obtains the normalized critical current:
\begin{eqnarray}
J_c= \frac{I_c ( H,\bm r_0)}{ g_cW} =\sqrt{A^2+B^2}
\label{Jc}
\end{eqnarray}

It is worth noting that $\varphi_H(y)$ is an odd function relative to the strip middle, whereas for a general vortex position $\varphi_v(y)$ is neither odd nor even unless $y_0=1/2$. In the latter case $\varphi_v(y)$ of Eq.\,(\ref{phi-vort}) is also odd relative to the strip middle; as a result $A=0$ and $J_c=|B|$.

It is readily seen that the critical current (\ref{Jc}) can also be written as
\begin{eqnarray}
J_c= \left | \int_0^1e^{i ( \varphi_H+\varphi_v )}dy \right |.
\label{Jc1}
\end{eqnarray}
In some situations  this form of $J_c $ is more convenient.

  Below, we consider a few cases of interest.

\subsubsection{No  vortex is present}

\begin{figure}[h]
\begin{center}
 \includegraphics[width=7cm] {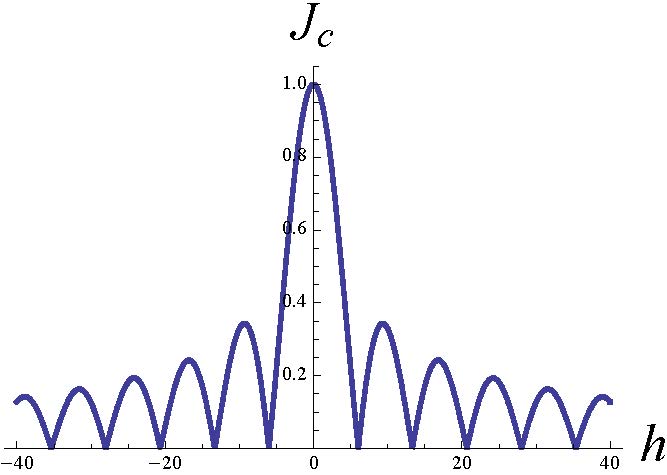}
\caption{ (Color online) The normalized Josephson critical current $J_c=I_c/g_cW$      vs $h=4HW^2/\phi_0$  in the vortex absence.}
\label{f3}
\end{center}
\end{figure}

The normalized critical current $J_c=I_c/g_cW$ evaluated with the help of Eqs.\,(\ref{I}) and  (\ref{Jc}) is shown in Fig.\,\ref{f3} versus reduced field $h=4HW^2/\phi_0$.  As expected,  in vortex absence, $J_c(h)$ is   symmetric with respect to $h=0$ at which $J_c $ reaches maximum,   the behavior characteristic of 0-type junctions.

\subsubsection{Vortex is far from the junction, $x_0>2$}

  In this situation the vortex contribution to the phase difference at the junction is a $y$ independent constant given in Eq.\,(\ref{C}). Then Eq.\,(\ref{Jc1}) shows that the vortex has no effect on the the pattern $J_c(h)$.
 We thus conclude that the vortex at a distance $x_0>2W$ does not affect the Fraunhofer pattern of the junction.

 \begin{figure}[h]
\begin{center}
 \includegraphics[width=7cm] {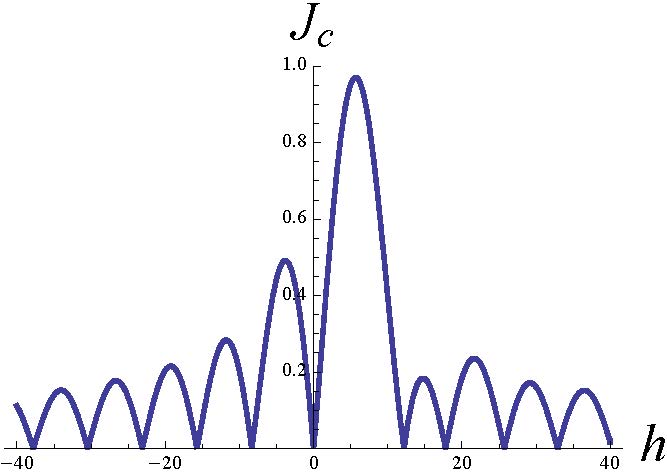}
\caption{ (Color online) The normalized Josephson critical current $J_c(h)$  for the vortex situated close to the junction: $x_0=0.17485$, $y_0=0.5$.}
\label{f4}
\end{center}
\end{figure}

%%%%%%%%
 \subsubsection{Zero-field $J_c(0,\bm r_0)$}
 %%%%%%%

 One of the relevant characteristics of the  pattern $J_c(h)$ is the value of zero-field critical current $J_c(0)$. In particular,  $J_c(0)=0$  signals a qualitative difference of the junction from the  '0-type'. To find $J_c(0)$  we start with Eq.\,(\ref{Jc1}) with $\varphi_H=0$.  The vortex factor $e^{i  \varphi_v }$ can be transformed using the logarithmic form of the inverse tangent in Eq.\,(\ref{phi-vort}) for $ \varphi_v  $:
 \begin{eqnarray}
 \varphi_v  &=&-2 \tan^{-1}u= -i \ln\frac{i+u}{i-u}\,,\label{fv}\\
  u&=&\frac{\cos \pi y-\cosh \pi x_0\,\cos \pi y_0}{\sinh \pi x_0\,\sin \pi y_0}.
\label{u}
\end{eqnarray}
We thus obtain
 \begin{eqnarray}
J_c(0)= \left | \int_0^1 \frac{i+u}{i-u}\,dy \right |.
\label{Jc2}
\end{eqnarray}
One can go here to  integration over $u$:
 \begin{eqnarray}
  u&=& C\cos \pi y-D\,,\nonumber\\
  C&=&\frac{1}{\sinh  \pi x_0\,\sin \pi y_0}\,,\quad D=\frac{\cosh \pi x_0\,\cos \pi y_0}{\sinh  \pi x_0\,\sin \pi y_0}\,,\qquad
\label{AB}
\end{eqnarray}
and obtain
 \begin{eqnarray}
J_c(0,\bm r_0)&=&\frac{1}{\pi} \left | \int_{u_1}^{u_2}\frac{du}{\sqrt{(u-u_1)(u-u_2)}}\,\frac{i+u}{i-u}  \right |,\nonumber\\
  u_1&=&-C-D\,,\qquad u_2=C-D\,.
\label{Jc3}
\end{eqnarray}

%%%%%%%%
 \subsubsection{The vortex in the strip middle}
 %%%%%%%

It is shown in this section that a vortex at some positions at the strip middle has an exclusive property to cause a shift in the pattern $J_c(h)$ so that instead of maximum at $h=0$, $J_c(0)$ is zero, the feature commonly ascribed to $(0,\pi)$ junctions.

\begin{figure}[h]
\begin{center}
 \includegraphics[width=9cm] {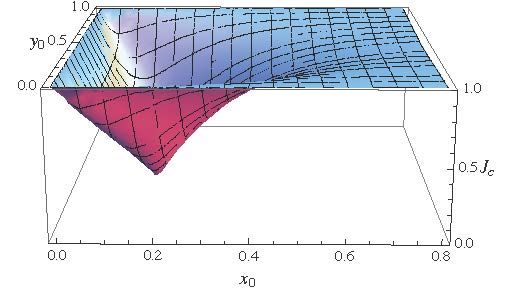}
\caption{(Color online) The   normalized zero-field Josephson critical currents $J_c(0,x_0,y_0)$.  The sharp minimum corresponds to $J_c(0,0.175,0.5)=0$. It is seen that this zero is isolated and no other zeros of $J_c(0)$ exist for a single vortex at this point. }
\label{4a}
\end{center}
\end{figure}

To find these positions we   note that for $y_0=1/2$, $C=1/\sinh \pi x_0$ and $D=0$. The integral in Eq.\,(\ref{Jc3}) then takes the form
 \begin{eqnarray}
{\cal J} =  \int_{-A}^{A}\frac{du}{\sqrt{u^2-C^2}}\,\frac{i+u}{i-u} =i\int_0^\pi\frac{i+C\cos v}{i-C\cos v} dv.\qquad
 \label{calJ}
\end{eqnarray}
where the substitution $u=C\cos v$ has been used. The last integral here can be written as $\int_0^{2\pi}dv/2  $ since only $\cos v$ enters the integrand. Further substitution $z=e^{iv}$ transforms the integral to a contour integral over the unit circle   in the complex plane $z$:
 \begin{eqnarray}
{\cal J} = \frac{1}{2}\oint  \frac{dz}{  z }\,\frac{z^2+2iz/C+1}{z^2-2iz/C+1} \,.
 \label{calJ1}
\end{eqnarray}
The product of the roots of $z^2-2iz/C+1=0$ is unity, hence only one of them is inside the unit circle. Then one readily obtains
 \begin{eqnarray}
{\cal J} =-i\pi\left(1-\frac{2}{\sqrt{1+C^2}}\right) = -i\pi(1-2\tanh\pi x_0) .\qquad
 \label{calJ2}
\end{eqnarray}
Thus, the zero-field critical current for a vortex at $(x_0,1/2)$ is:\cite{Clem}
 \begin{eqnarray}
J_c(0,x_0,1/2) =|1-2\tanh(\pi x_0)|.
 \label{Jcrit}
\end{eqnarray}
It is seen that $J_c(0,x_0,1/2) $ has only one root $x_0\approx 0.175$.  At $x_0=0$ and approximately for $x_0>2$ $J_c(0,x_0,1/2)=1 $ in agreement with the earlier conclusion that the far-away vortex does not matter for $J_c(h)$. Moreover, the point $(0.175,0.5)$ of the plane $(x_0,y_0)$ is the only one (for a single vortex) where $J_c(0,x_0,y_0)=0 $. This is seen in Fig.\,\ref{4a} where $J_c(0,x_0,y_0)  $ is evaluated numerically using Eq.\,(\ref{Jc3}).

\begin{figure}[t]
\begin{center}
 \includegraphics[width=7cm] {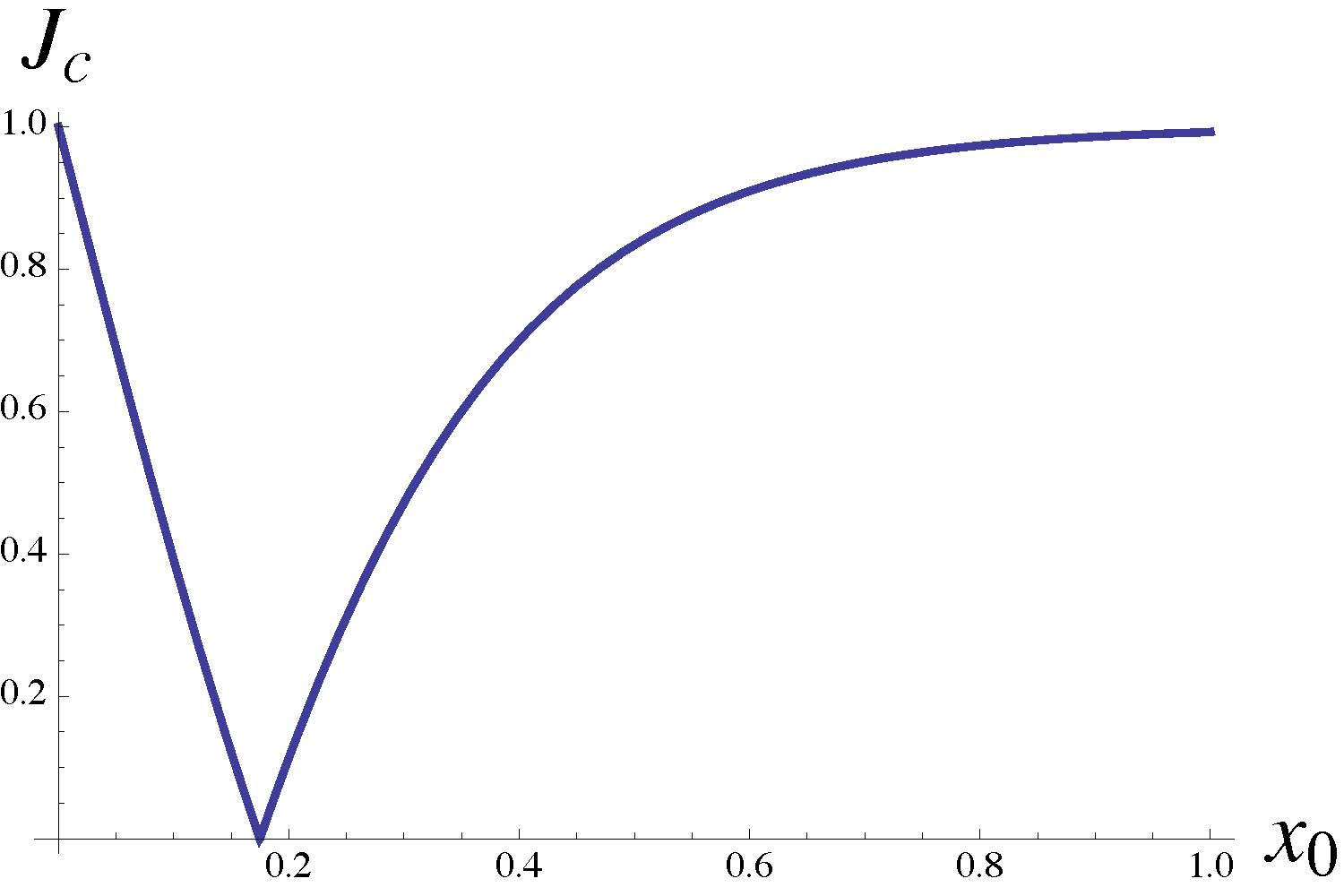}
\includegraphics[width=7cm] {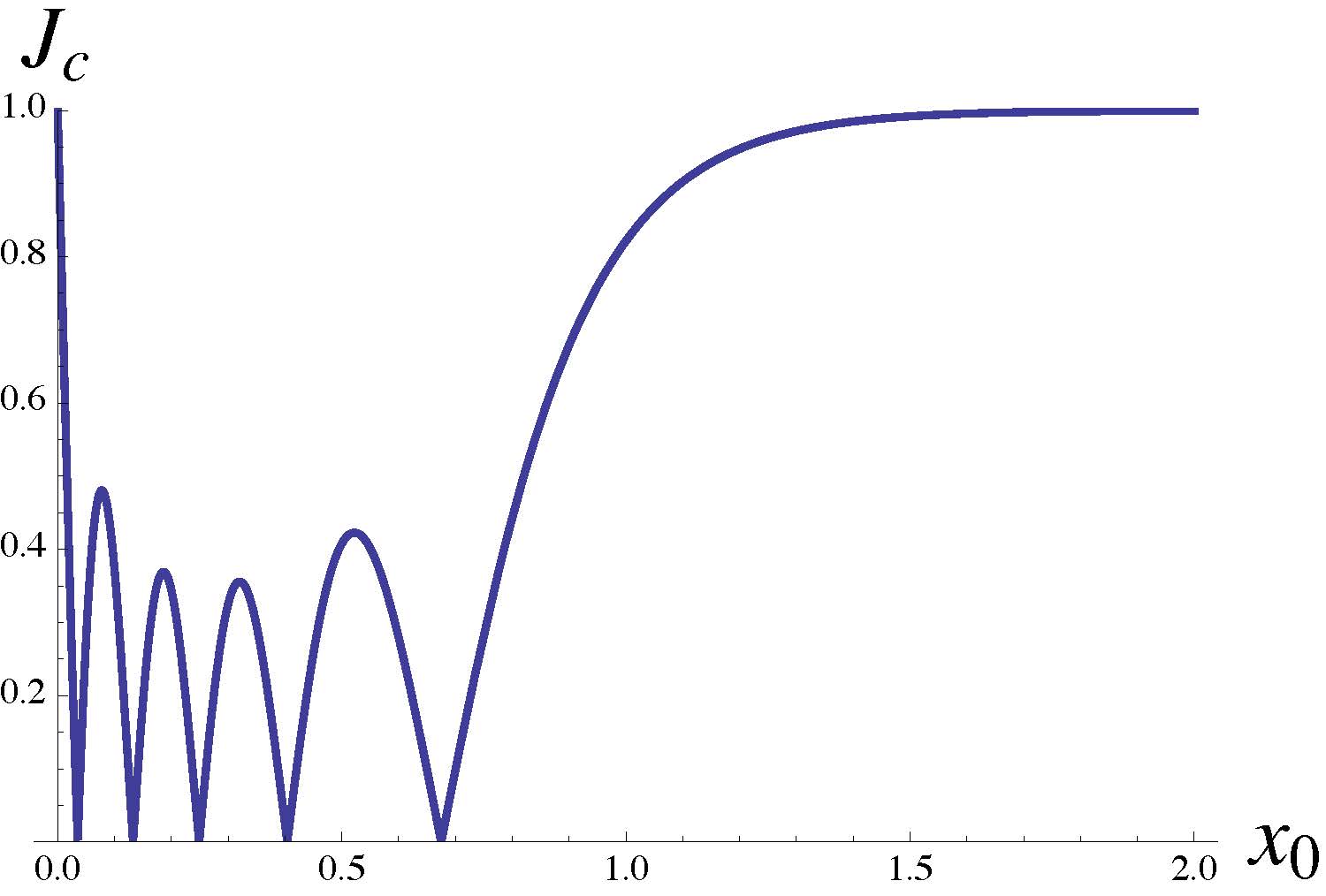}
\includegraphics[width=7cm] {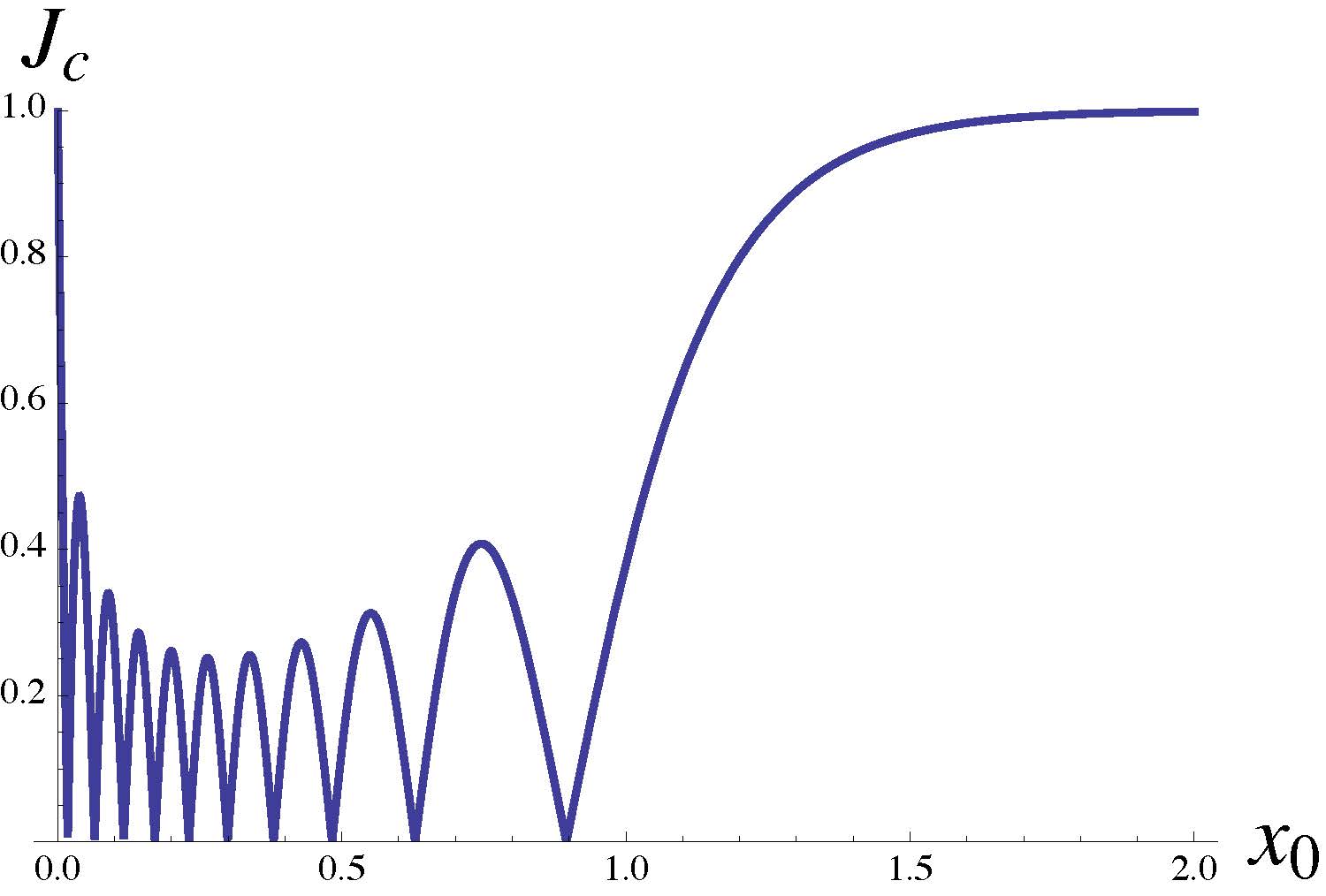}
\caption{(Color online) The   normalized zero-field Josephson critical currents $J_c(0)$  for  vortices   in the strip middle  $y_0=0.5$ as a function of $x_0$: the upper panel is for $N=1$, the middle panel  $N=5$, and the lowest panel $N=10$.  Roughly, the intervals $\Delta x_0 \propto 1/(N+1-n)$ where $n$ is the number of the zero counted from $x_0=0$.  }
\label{f6}
\end{center}
\end{figure}

If $N$ vortices are trapped at the same point $\bm r_0=(x_0,y_0)$, the vortex phase (\ref{fv}) acquires a factor $N$. As a result one has to replace the factor $(i+u)/(i-u)$ in Eqs.\,(\ref{Jc2}), (\ref{Jc3}),    (\ref{calJ}) with $(i+u)^N/(i-u)^N$. In turn, this leads to a pole of the order $N$ inside the unit circle in integration over $z$. In principle, one can proceed with analytical evaluation, but the result is increasingly cumbersome with increasing $N$. We resort then to numerical evaluation of $J_c(0,x_0,1/2)$  examples of which are shown in Fig.\,\ref{f6}. It is seen that the number of positions $x_0$ for which the Fraunhofer pattern has zero at $h=0$ is equal to the number of vortices trapped at $x_0$. The density of these points also increases with $N$, so that for large number of vortices trapped, nearly any place $x_0$ of the trap in the interval  $0<x_0\lesssim 2$ will make the  pattern $J_c(h,x_0,1/2)$ to have near-zero at $h=0$. The upper bound of this interval is related to the fact that for $ x_0\gtrsim 2$ the vortex effects upon the Fraunhofer pattern vanish  and $J_c(0) $ approaches unity  exponentially as is seen from Eq.\,(\ref{Jcrit}).

 \begin{figure}[t
]
\begin{center}
 \includegraphics[width=7cm] {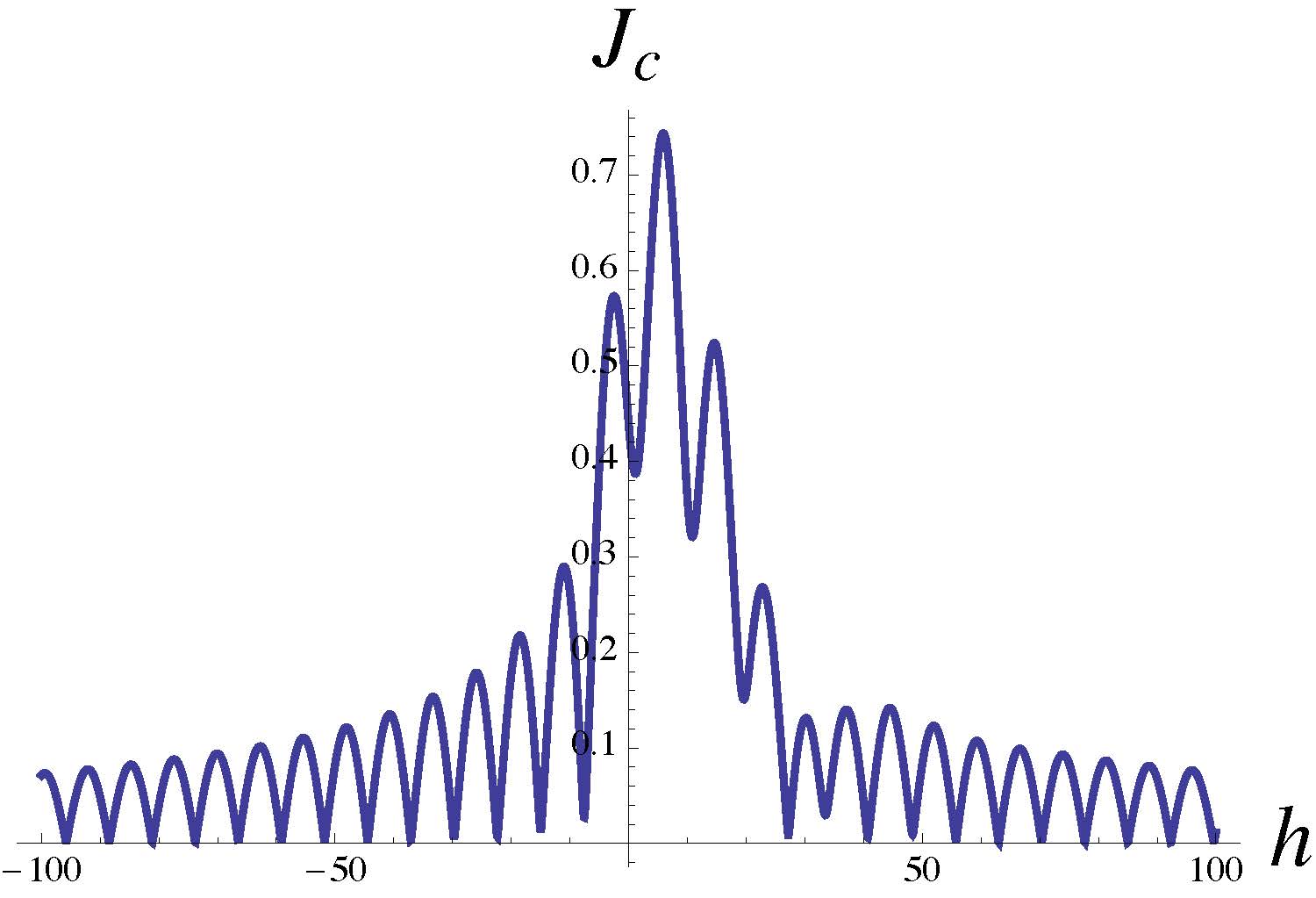}
\includegraphics[width=7cm] {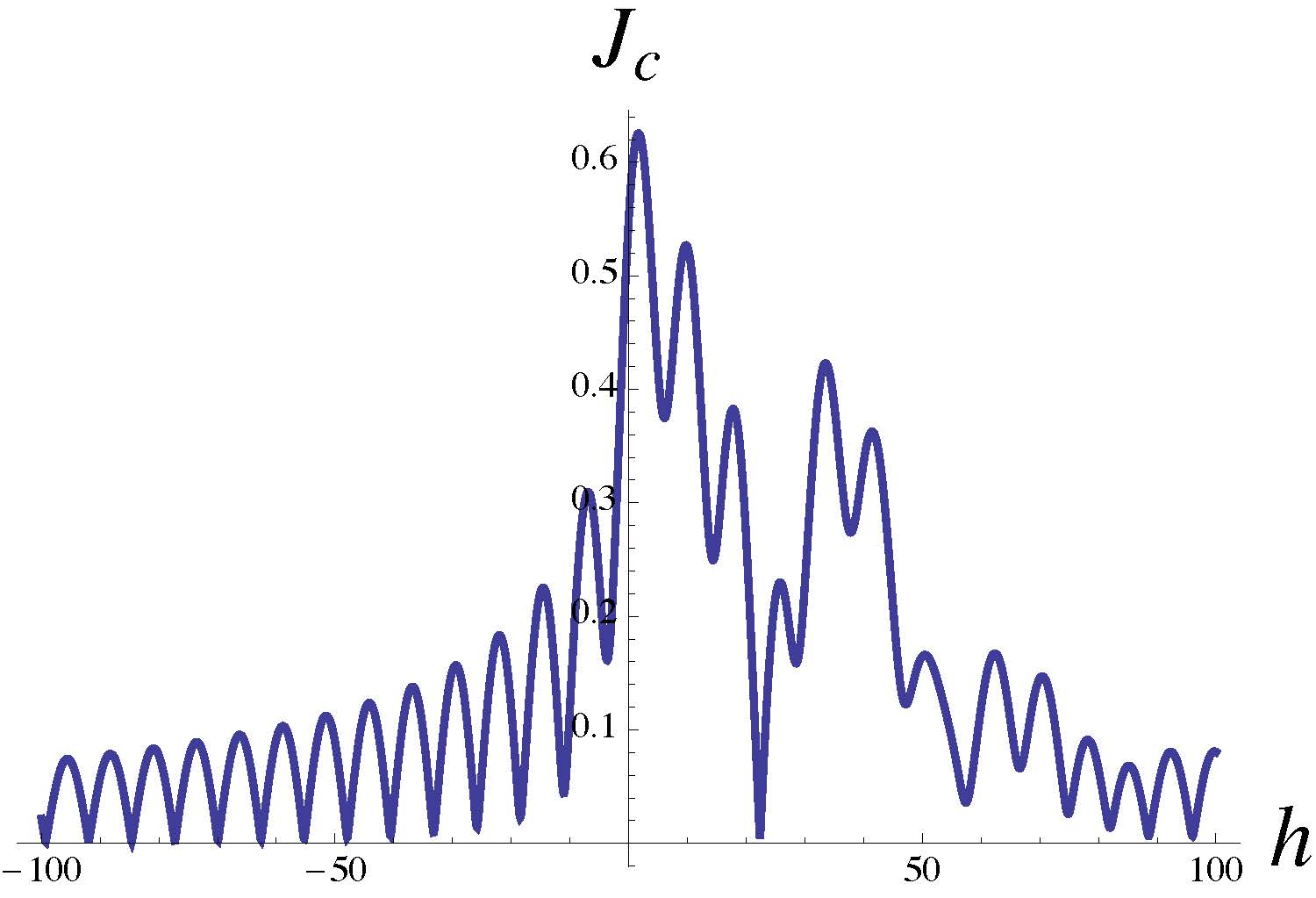}
\includegraphics[width=7cm] {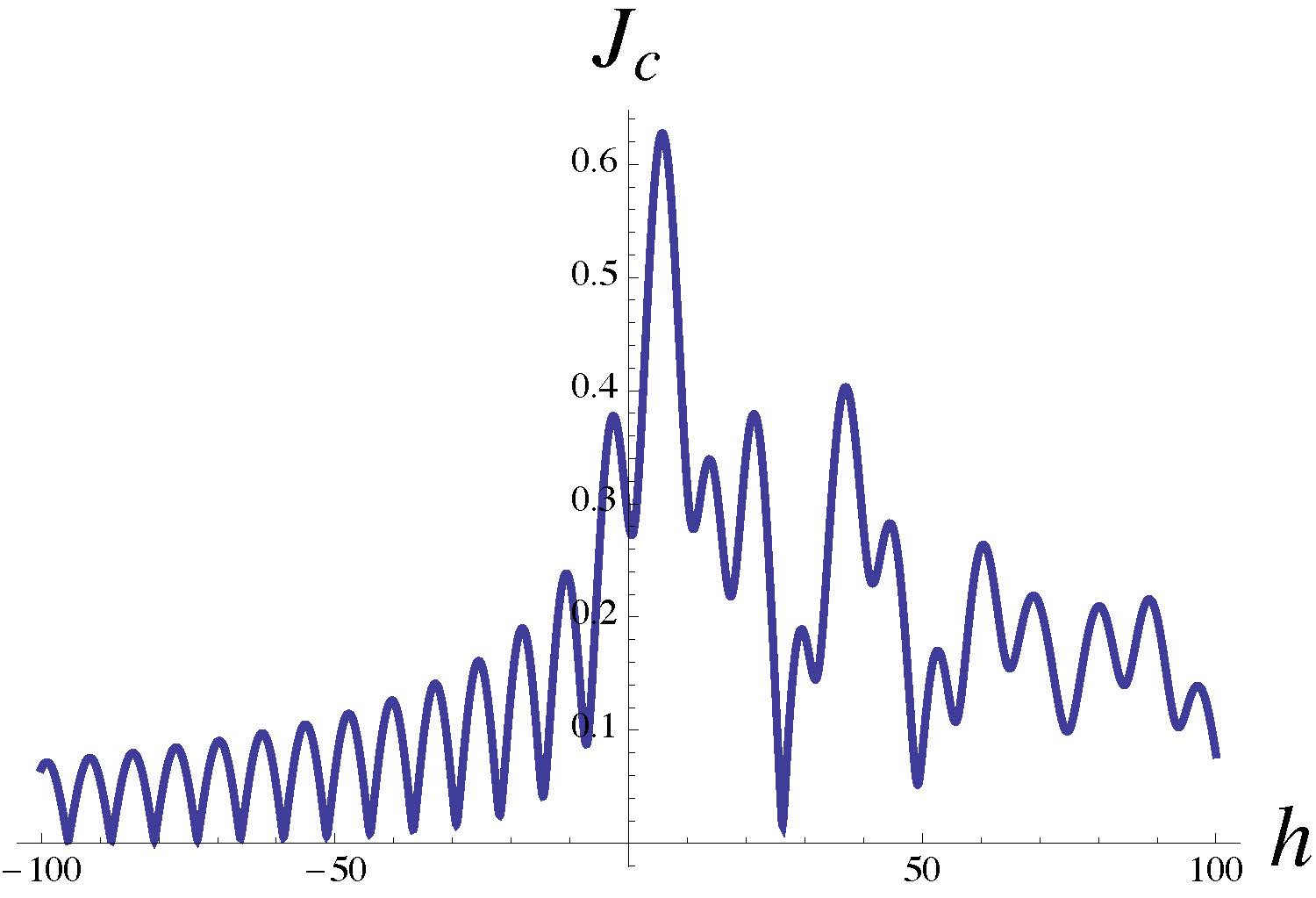}
\caption{(Color online) The upper panel: normalized Josephson critical currents $J_c(h)$  for  a vortex at    $x_0=0.1$,     $y_0=0.3$. The middle panel: the same for 3 vortices trapped at the same location.  The lowest panel: the same for 5 vortices trapped at the same location. }
\label{f7}
\end{center}
\end{figure}
 \begin{figure}[t]
\begin{center}
 \includegraphics[width=7cm] {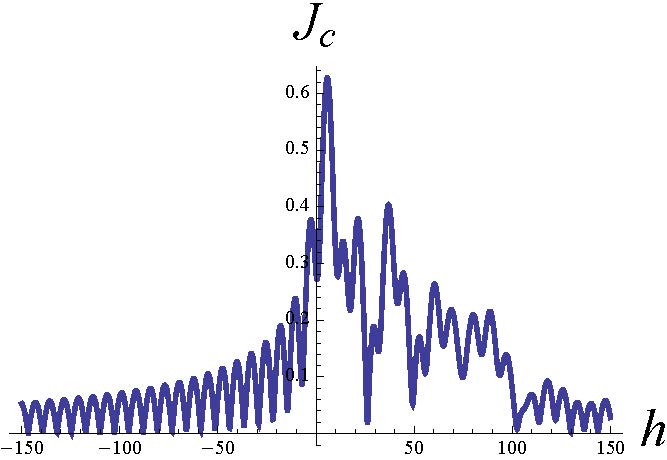}
\includegraphics[width=7cm] {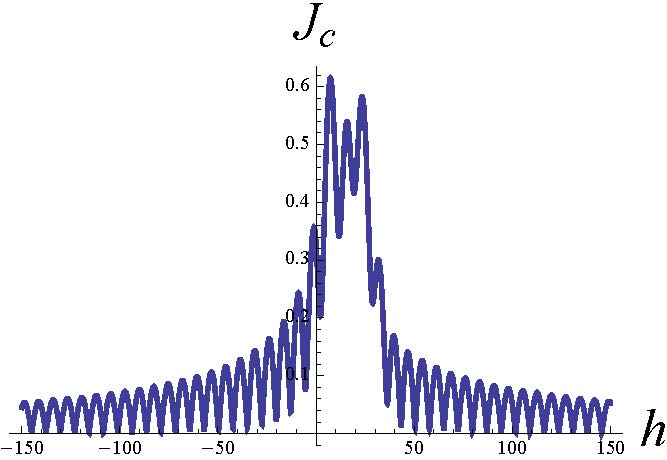}
\includegraphics[width=7cm] {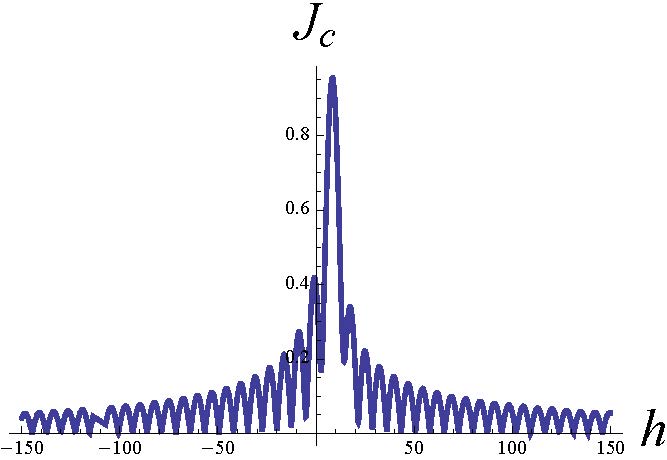}
\caption{(Color online) The upper panel: normalized Josephson critical currents $J_c(h)$  for five vortices    $x_0=0.1$,     $y_0=0.3$. The middle panel: the same for $\bm r_0=(0.3,0.3)$.    The lowest panel:  5 vortices trapped at $\bm r_0=(0.5,0.3)$.   }
\label{f8}
\end{center}
\end{figure}

%%%%%%%
\subsubsection{Arbitrary position of a near-by vortex}
%%%%%%

 The upper panel of Fig.\,\ref{f7} shows $J_c(h)$ for a sigle vortex at $x_0=0.1$, $ y_0=0.3$. Characteristic features of this $J_c(h)$ are the presence of non-zero minima and a strong asymmetry of the pattern relative to $h\to -h$. The effect of a vortex is strongest on the side of positive $h$. This is seen better yet if   two vortices are trapped at the same position $\bm r_0=(0.1,0.3)$, the middle panel, or five shown in the lowest panel. Note that the pattern at $h<0$ is well ordered with a repetition step $\Delta h\approx 7.1$ which corresponds to  $\Delta H\approx 1.8\phi_0/W^2$ as should be for a pattern caused exclusively by the applied field $H$.\cite{Beasley,Maayan,Clem}

  One can see  in Fig.\,\ref{f8}
  an example of $J_c(h)$ for 5 vortices trapped at the same transverse coordinate $y_0=0.3$ but at increasing separations $x_0=0.1,0.3,0.5$. We have chosen a broader domain $|h|<150$ to show that vortex effects on the right side of the pattern persist  up to a large value of $h$, which however decreases with increasing separation.

 If the vortex approaches the strip edges $y_0=0 $ or 1, $J_c(h)$ approaches the pattern shown in Fig.\,\ref{f3} for no vortices. As  argued in Ref.\,\onlinecite{last}, in this case the vortex causes the junction phase difference to acquire an extra $\pi$, which does not change the tunneling current, but affects the junction energy.

%%%%%%%%
\section{Discussion.}
We have  shown that a vortex at one of the banks of the plane thin-film Josephson junction distorts the pattern of the field dependent critical current $J_c(h)$ in a strongly asymmetric way: as is seen in Figs.\,\ref{f4}, \ref{f7}, \ref{f8}, the distortion at the side $h>0$ for a vortex is strong, whereas for $h<0$ it is weak and more regular (for antivortex the picture flips).  Actually, this asymmetry is seen in experiment.\cite{Krasnov}

We also show that the vortex effect upon Fraunhofer pattern $J_c(h)$ disappears exponentially when  the junction-vortex separation $x_0\gtrsim 2W$ with the length scale $W/\pi$. This, however, does not mean that the junction ``does not feel" the far-away vortex; as Eq.\,(\ref{C}) shows, the junction phase difference acquires a constant addition dependent on the transverse vortex coordinate $y_0$.\cite{last}   Hence, the junction energy influenced by the vortex for all junction-vortex separations.

In principle, effects discussed here open possibilities to manipulate properties of Josephson junctions by trapping vortices in junction banks.
We identified a number of vortex positions $(x_0,1/2)$ for which  the zero-field critical current $J_c(0)$ turns   zero. Hence, by measuring  $J_c(0)$ one can say whether or not one of these positions  $(x_0,1/2)$ is occupied by a vortex, an interesting possibility for applications.

  Our calculations are valid for sufficiently thin and narrow superconducting  strips for which the condition $W\ll \Lambda$, the Pearl length, is satisfied. This condition allows us to disregard the magnetic energy of supercurrents relative to their kinetic energy. For other types of junctions (e.g., made of thick overlapping films) our solutions {\it per se} do not apply.  \\

 The authors are grateful to  J. Kirtley, I. Sochnikov, A. Ustimov,  J. Mannhart, and S. Lin for helpful  discussions. This work was supported by the U.S. Department of Energy, Office of Science, Basic Energy Sciences, Materials Science and Engineering Division. The work was done at the Ames Laboratory, which is operated for the U.S. DOE by Iowa State University under contract  DE-AC02-07CH11358.

\end{document}